\documentclass[twocolumn,superscriptaddress,showpacs,floatfix,preprintnumbers, nofootinbib,hyperref]{revtex4} 
\usepackage{graphicx}%  Default Latex 2eps package for embedding figures
\usepackage{rotating}
\usepackage{epsfig}
\usepackage{bm}
\usepackage[usenames,dvipsnames]{xcolor}
\usepackage{color}

\def\he4{$^4$He}
\def\h2{$^2$H}

\newcommand{\lesssim}{\,\rlap{\lower3.7pt\hbox{$\mathchar\sim$}}
\raise1pt\hbox{$<$}\,}

\begin{document}

\preprint{ IPPP/15/08; DCPT/15/16}

\title{{Self-induced flavor instabilities of a dense neutrino stream in  a two-dimensional model}}

\author{Alessandro Mirizzi} 
\affiliation{II Institut f\"ur Theoretische Physik, Universit\"at Hamburg, Luruper Chaussee 149, 22761 Hamburg, Germany.}
\affiliation{Dipartimento Interateneo di Fisica ``Michelangelo Merlin'', Via Amendola 173, 70126 Bari.}
\affiliation{Istituto Nazionale di Fisica Nucleare - Sezione di Bari, Via Amendola 173, 70126 Bari, Italy}
\author{Gianpiero Mangano}
\affiliation{Istituto Nazionale di Fisica Nucleare - Sezione di Napoli, Complesso Universitario di Monte S. Angelo, I-80126 Napoli, Italy}
 \author{Ninetta Saviano} 
\affiliation{Institute for Particle Physics Phenomenology, Department of Physics,
Durham University,\\ Durham DH1 3LE, United Kingdom}

%\date{\today}

\begin{abstract}
We consider a  
simplified model for self-induced flavor conversions of a  dense neutrino gas in two  dimensions,  showing new solutions that 
spontaneously
break the spatial symmetries of the initial conditions.
As a result of the symmetry breaking induced by   the neutrino-neutrino interactions,
 the coherent behavior of the neutrino gas becomes unstable. This instability produces
large spatial variations in the flavor content of the ensemble.  
Furthermore, it also leads to the creation of domains of different net lepton number flux. 
The transition of the neutrino gas from a coherent to incoherent behavior  shows an intriguing  analogy  with  a streaming flow changing from  laminar to  turbulent regime. These finding would be relevant for the self-induced conversions of neutrinos streaming-off a supernova core. 
\end{abstract}


\maketitle

\section{Introduction}
%%%%%%%%%%%%%%%%%%%%%%%%%%%%%%%%%%%%%%%%%%%%%%%%%%%%%
Neutrino-neutrino interactions in dense neutrino gases  provide a 
refractive term leading to a non-linear feedback in the flavor evolution,
when the neutrino interaction potential $\mu \sim \sqrt{2}G_F n_\nu$    is comparable or larger  than the neutrino vacuum oscillation
 frequency $\omega=\Delta m^2/2E$.
This effect can lead 
to a collective behavior of the dense neutrino ensemble (see~\cite{Duan:2010bg} for a recent review).
Remarkably, almost a decade ago  it was realized   that neutrino-neutrino
interactions dominate the  flavor evolution in the deepest core--collapse supernova (SN) regions producing  self-induced flavor
 conversions~\cite{Duan:2006an,Hannestad:2006nj,Fogli:2007bk}.
The most important observable consequence of this type of flavor transitions would be the   
 swap of the SN $\nu_e$ and $\bar\nu_e$ spectra with that of
non-electron flavor $\nu_x$ and $\bar\nu_x$ in certain energy intervals~\cite{Dasgupta:2009mg}.  

Characterizing the SN neutrino flavor dynamics
amounts to follow the spatial evolution of the neutrino fluxes. For a stationary
neutrino emission,
the kinetic equations of 
the  $\nu$ space-dependent occupation numbers  $\varrho_{{\bf p}, {\bf x}}$
with momentum ${\bf p}$ at position ${\bf x}$ are~\cite{Sigl:1992fn,Strack:2005ux} 
%..........................................................
\begin{eqnarray}
&& {\bf v}_{\bf p} \cdot \nabla_{\bf x}\, \varrho_{{\bf p}, {\bf x}} 
 = - i [\Omega_{{\bf p}, {\bf x}}, \varrho_{{\bf p}, {\bf x}}]
\,\ ,
\label{eq:eom}
\end{eqnarray}
%........................................................
with the Liouville operator in the left-hand side.  Neglecting external forces
and an explicit time dependence of the occupation numbers, this operator  represents the drift term proportional to the neutrino velocity
${\bf v_p}$, due to particle free streaming. 
On the right-hand-side the matrix $\Omega_{{\bf p}, {\bf x}}$ is the full Hamiltonian containing the
vacuum, matter and self-interaction terms. 
In particular, in non-isotropic neutrino gases, like the case of neutrinos streaming-off a SN core, 
the neutrino-neutrino interaction term 
contains {\it multi-angle} effects since
the current-current nature of the low-energy weak interactions introduces
an angle dependent term  $(1-{\bf v}_{\bf p} \cdot {\bf v}_{\bf q})$ between two interacting neutrino modes~\cite{Qian:1994wh,Duan:2006an}.
This term produces a net current so that test neutrinos moving in different
directions would experience different refractive index. This in some cases   challenges the collective behavior of 
the flavor evolution observed in an isotropic case, leading to   {\it flavor decoherence}~\cite{Raffelt:2007yz,EstebanPretel:2007ec,Sawyer:2008zs}. 
Multi-angle effects can also lead to a trajectory-dependent matter term, which if  strong enough  suppresses the self-induced
conversions~\cite{Chakraborty:2011nf,Sarikas:2011am}. 

The multi-angle  flavor evolution described by the partial differential
equations~(\ref{eq:eom})  has never been solved till now in its full complexity.
Instead,  numerical approaches have been typically based
on the so-called  
\emph{``bulb model''}~\cite{Duan:2006an,Fogli:2007bk,EstebanPretel:2007ec}, where it is assumed a spherical symmetry about the center of the SN and azimuthal
symmetry about any radial direction. In this limit Eq.~(\ref{eq:eom}) reduces to an ordinary differential equation problem, projecting the evolution along the radial direction, i.e. 
$ {\bf v_p}\cdot \nabla_{\bf x} \to  v_r\,  d/dr$.

Attempts to go beyond the bulb model have been proposed. For example, in~\cite{Dasgupta:2008cu} it was shown that  assuming that the neutrino ensemble displays self-maintained coherence, 
the problem for generic geometries can be reduced to a  one-dimensional case along the streamlines of the overall neutrino flux. 
However, the existence of a self-consistent  coherent solution does not imply its stability. 
Indeed, it has been recently realized  that instabilities may grow once one relaxes some symmetries of the bulb model, 
since neutrino-neutrino interactions can lead to \emph{spontaneous symmetry breaking} effects.
In~\cite{Sawyer:2008zs} it was shown that perturbing the  symmetry of the  initial neutrino angular distributions one would find
a speed-up in the  flavor exchange.
Moreover, removing the assumption of axial symmetry in the $\nu$ multi-angle term, a  multi-azimuthal-angle instability
has been discovered, even  assuming a perfect spherically symmetric $\nu$ 
emission~\cite{Raffelt:2013rqa,Raffelt:2013isa,Duan:2013kba,Mirizzi:2013rla,Mirizzi:2013wda}.
Furthermore, also space and time homogeneity can be broken in a dense neutrino gas, so that it is not guaranteed that
a quasi-stationary neutrino emission would lead to a stationary solution~\cite{Mangano:2014zda}.  
Finally, with a simple  toy model it has been recently shown, by means of  a stability analysis of  the linearized equations of motion,
that self-induced oscillations can spontaneously break  spatial symmetries~\cite{Duan:2014gfa}.
All these findings suggest that the validity of the bulb model should be critically reconsidered and that a self-consistent solution of the SN neutrino flavor evolution can only be achieved by solving the complete \emph{multi-dimensional} problem of Eq.~(\ref{eq:eom}).

As a further step in clarifying this issue we consider here the two-dimensional toy  model discussed in~\cite{Duan:2014gfa}, i.e. 
 monochromatic neutrino streams emitted in a stationary way in two directions from an infinite boundary plane at $z=0$ with periodic conditions on $x$ and translation invariance along the $y$ direction.  
We assume a small perturbation for   the initial symmetries of the flavor content in both the two emission modes and along the boundary in the $x$ direction.
Despite of its simplicity, this model with perturbed symmetries exhibits a rich phenomenology.  
Indeed,  solving numerically the flavor evolution  we find that the initial small perturbations are amplified by neutrino interactions, leading to  non-trivial two-dimensional structures in the oscillation pattern that exhibits large space fluctuations. Any particular solution which is required to satisfy initial symmetries at the boundary is thus, unstable, confirming the results of~\cite{Duan:2014gfa}. We will show that the 
neutrino flux transition  from a coherent to incoherent behavior resembles quite closely the transition from a laminar to  turbulent regime of a streaming flow. This issue will be studied in more details elsewhere.
Finally, also the lepton number flux would develop a domain structure.
 
\section{A two-dimensional model}
%%%%%%%%%%%%%%%%%%%%%%%%%%%%%%%%%%%%%%%%%%%%%%%%%%%%%%%%%%%%%%%%
As  in~\cite{Duan:2014gfa}, neutrinos are  emitted from an infinite plane at $z=0$,
in two directions
%..........................................................
\begin{equation}
{\hat {\bf v}}_\zeta = (v_\zeta,0,v_z) \,\ \,\ \,\ \,\ \,\ (\zeta = L,R) \,\ ,
\end{equation}
%........................................................... 
where $0 < v_z<1$ and $v_R=-v_L =\sqrt{1-v_z^2}$. The flavor evolution occurs
for $z >0$ in the two-dimensional plane  spanned by the $x$ and $z$ coordinates.
We consider a two-flavor $(\nu_e,\nu_x)$ neutrino ensemble. Expanding all quantities of  Eq.~(\ref{eq:eom})
in terms of Pauli matrices $\sigma$, one gets the equations for the $L$  mode
%..........................................................
\begin{eqnarray}
{\hat {\bf v}}_L \cdot \nabla_{\bf x} {\sf P}_L(x,z)  &=& [+\omega {\sf B}  
 + \mu  {\sf D}_R(x,z)] \times {\sf P}_L(x,z) \, , \nonumber   \\
 {\hat {\bf v}}_L \cdot \nabla_{\bf x} {\overline{\sf P}}_L(x,z)  &=& [-\omega {\sf B}  
 + \mu  {\sf D}_R(x,z)] \times  {\overline{\sf P}}_L(x,z)   \, ,
%\nonumber  \\
%{\hat v}_R \cdot \nabla_{\zeta} {\sf P}_R(x,z)  &=& [+\omega {\sf B}  
% + \mu  {\sf D}_L(x,z)] \times {\sf P}_R(x,z) , \nonumber \\
%  {\hat v}_R \cdot \nabla_{\zeta} {\overline{\sf P}}_R(x,z)  &=& [-\omega {\sf B}  
% + \mu  {\sf D}_L(x,z)] \times  {\overline{\sf P}}_R(x,z) , 
\label{eq:evolspace}
\end{eqnarray}
with   $\omega=\Delta m^2/2 E$ the vacuum oscillation frequency.
%.............................................................
The equations from the $R$ mode can
be obtained from the previous ones using the $L \leftrightarrow  R$ symmetry.

The sans-serif in Eq.~(\ref{eq:evolspace}) indicates three-dimensional vectors in flavor space, e.g. 
${\sf B} = (B^1, B^2,B^3)$. 
 The differential operator at left-hand-side reads
%.....................................................
\begin{equation}
{\hat {\bf v}}_{L} \cdot \nabla_{\bf x} = v_{L} \partial_x  
+ v_z \partial_z  \,\ .
\end{equation}
%......................................................
The ${\sf P}_{L,R}$ ($\overline{\sf P}_{L,R}$) functions are the neutrino (antineutrino) polarization vectors in flavor space
for the $L,R$ modes, defined from the occupation numbers as in~\cite{EstebanPretel:2007ec}, i.e.
$\varrho_{L,R}=1/2[F_{\nu_e}+F_{\nu_x} + (F_{\bar\nu_e}-F_{\bar\nu_x}) {\sf P}_{L,R}\cdot \sigma]$ and
${\overline\varrho}_{L,R}=1/2[F_{\bar\nu_e}+F_{\bar\nu_x} + (F_{\bar\nu_e}-F_{\bar\nu_x}) {\overline{\sf P}}_{L,R}\cdot\sigma]$, where
$F_{\nu}$ are the total neutrino fluxes of the different species.
 We define as usual ${\sf D}_{L,R} =  {\sf P}_{L,R} - \overline{\sf P}_{L,R}$.
 The unit vector ${\sf B}$ points  in the mass eigenstate direction in flavor space, such 
that ${\sf B}\cdot{\sf e}_3=-\cos \theta$, where $\theta$ is the vacuum mixing angle.
For simplicity we neglect a possible matter effect, assuming that its only role would be to reduce
the effective in-medium mixing angle, $\theta \ll 1$~\cite{Hannestad:2006nj}.  
Finally, the neutrino self-interaction term is given by~\cite{EstebanPretel:2007ec}
%..........................................................
\begin{equation}
\mu = \sqrt{2} G_F [F^0_{\bar\nu_e}-F^0_{\bar\nu_x}]
(1-{\hat {\bf v}}_L \cdot {\hat {\bf v}}_R) \,\ ,
\end{equation}
%......................................................
expressed in terms of the antineutrino fluxes $F^0_{\bar\nu}$ at the boundary. 
Since in our model  the neutrino trajectories intersect at a fixed angle, the  previous
equation would correspond to
 the ``single-angle'' approximation~\cite{Raffelt:2007yz}. Therefore, multi-angle effects
 do not play a role in this case.
 
One can define a conserved ``lepton current'' 
${\textrm L}^\mu =({\textrm L}_0, {\bf L})$ whose components are
(see~\cite{Duan:2008fd})
%...................................................
\begin{eqnarray}
{\textrm L}_0 &=& {\sf D}_L \cdot{\sf B} +  {\sf D}_R \cdot{\sf B} \,\ , \label{eq:lepton0} \\
{\bf L} &=& {\hat {\bf v}}_L ({\sf D}_L \cdot{\sf B}) +  {\hat {\bf v}}_R ({\sf D}_R \cdot{\sf B}) \,\ ,
\label{eq:lepton}
\end{eqnarray}
%....................................................
where  ${\bf L}$ is a two-dimensional vector in the $(x,z)$ plane, and 
${\sf D}_{L,R} \cdot{\sf B} \simeq P^3_{L,R} -{\overline P}^3_{L,R}$.  In the following
we will normalize ${\textrm L}^\mu$
to the number of modes $N_{L,R}=2$.
It is straightforward to show from Eq.~(\ref{eq:evolspace}) that ${\bf L}$
satisfies a continuity equation 
%..................................
\begin{equation}
\partial_t {\textrm L}_0 + \nabla_{\bf x} \cdot {\bf L} =\nabla_{\bf x} \cdot {\bf L}=0 \,\ ,
\label{eq:leptcons}
\end{equation}
%.......................................
where first equality follows from the assumption of a  stationary solution.
 Eq.~(\ref{eq:leptcons}) generalizes the lepton-number conservation law 
 of the one dimensional case~\cite{Hannestad:2006nj}.

As first exploited in~\cite{Mangano:2014zda} (see also~\cite{Duan:2014gfa}) in the context of  multi-dimensional neutrino oscillations, 
 the partial differential equation problem like of Eq.~(\ref{eq:evolspace}) can be reduced to a tower of ordinary 
differential equations for the Fourier modes defined as
%...................................
\begin{equation}
{\sf P}_{L(R),k}(z) = \int_{-\infty}^{+\infty} dx \,\ {\sf P}_{L(R)}(x,z) e^{-i k x} \,\ ,
\end{equation}
%..................................
and similarly for the antineutrino polarization vectors ${\overline{\sf P}}_{L,R}$.
In the following, 
for simplicity we will consider a monochromatic perturbation in the $x$ direction for 
the   neutrino polarization vectors at the boundary at
$z=0$. Since we start with a pure flavor state, we assume ${P}^1_{L,R}(x,0)={P}^2_{L,R}(x,0)=0$ and
\begin{equation}
{P}^3_{L,R}(x,0) = \langle {P}^3_{L,R}(x,0) \rangle +\epsilon \cos(k_0 x) \,\ , 
\label{eq:translperturb}
\end{equation}
%.................................................
where this latter component of the polarization vector is proportional to the flavor content of the ensemble. 
The function $\langle {P}^3_{L,R}(x,0) \rangle $ indicates the unpertubed value of the polarization vectors, while 
  $k_0$ is the wave-number of the perturbation and  $\epsilon \ll \mu , \omega$ its small amplitude. 
It is easy to see that  in this case, only higher harmonics of the fundamental mode
with $k_n = n k_{0}$ are excited. Defining  ${\sf P}_{L,n} = k_0 {\sf P}_{L,k_n}/(2\pi)$,
from Eq.~(\ref{eq:evolspace}) one obtains  

%................................................................
\begin{eqnarray}
v_z \frac{d}{dz} {\sf P}_{L,n}(z) &=& -i u_L k_n {\sf P}_{L,n} + \omega {\sf B}\times {\sf P}_{L,n}  \nonumber \\
&+& \mu \sum_{j=-\infty}^{+\infty} {\sf D}_{R,{n-j}}\times {\sf P}_{L,j} \,\ .
\label{eq:eompert}
\end{eqnarray}
%.....................................................................
An analogous set of coupled ordinary differential equations  can be written for the $R$ mode and for the antineutrino polarization vectors.
It is enough to follow the evolution for  positive modes, $n \geq 0$, since the ${\sf P}_{L,R}(x,z)$ and ${\overline{\sf P}}_{L,R}(x,z)$ are real functions and therefore
\begin{equation}
{{\sf P}^{\ast}_{(L,R),n}}= {\sf P}_{(L,R),-n} \,\ .
\end{equation}
Once the evolution of the harmonic modes is obtained from Eq.~(\ref{eq:eompert}), the polarization vector in configuration space can be obtained by inverse Fourier transform.

%%%%%%%%%%%%%%%%%%%%%%%%%%%%%%%%%%%%%%%%%%%%%%%%%%%%%%%%%%%%%%
\begin{figure}[!t]\centering
\hspace{-1.8cm}
\includegraphics[angle=0,width=1.1\columnwidth]{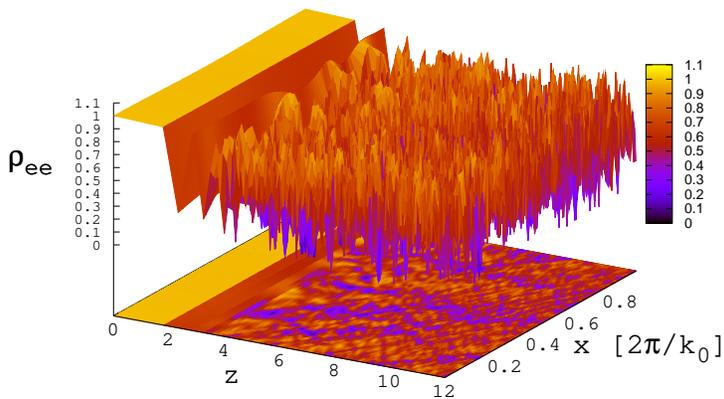}
%\vspace{-3.cm}
\caption{Three dimensional evolution of the $\nu_e$ flavor content $\varrho_{ee}$ in the $x$-$z$ plane, and its map  on the bottom plane.
\label{polar}}
\end{figure}
%%%%%%%%%%%%%%%%%%%%%%%%%%%%%%%%%%%%%%%%%%%%%%%%%%%%%%%%%%%%%

\section{Numerical examples}
%%%%%%%%%%%%%%%%%%%%%%%%%%%%%%%%%%%%%%%%%%%%%%%%%%%%%%%%%%
To illustrate the behavior of the self-induced flavor conversions in our two-dimensional toy model, we consider a gas initially 
composed by only $\nu_e$ and $\bar\nu_e$. We normalize the polarization vectors to the $\bar\nu_e$ number density and we 
assume an excess of $\nu_e$ over $\bar\nu_e$, i.e. $F^0_{\nu_e}/F^0_{\bar\nu_e} = 1 + \alpha$. 
If the translational symmetry is assumed, i.e. $\epsilon=0$ in Eq.~(\ref{eq:translperturb}), 
and the $L$ and $R$ modes are prepared identically, it is well known that the system is stable in normal mass hierarchy
($\Delta m^2 >0$), while in inverted mass hierarchy  ($\Delta m^2 <0$) it exhibits a bimodal instability and    behaves as a \emph{flavor pendulum}, leading
to periodic  pair conversions  $\nu_e \bar\nu_e \leftrightarrow \nu_x \bar\nu_x$, that conserve the lepton number 
${\textrm L}_0 =\alpha$ of Eq.~(\ref{eq:lepton0})~\cite{Hannestad:2006nj}. 
More recently, it has been shown that if the $L \leftrightarrow  R$ symmetry is perturbed the system becomes unstable also in normal hierarchy, with a similar pendulum behavior~\cite{Raffelt:2013isa}. 
Our further step is to perturb also the translational symmetry at the boundary. 

We fix in Eq.~(\ref{eq:evolspace}) $\mu=10$, $\omega=1$, $\theta=10^{-3}$, and we consider the normal mass hierarchy case (the result would be similar for the inverted mass hierarchy).
We take as asymmetry parameter $\alpha=0.3$.
We assume $v_R=v_z=\sqrt{2}/2$.
Perturbation in  the $L \leftrightarrow  R$ symmetry are introduced by 
 a $1\%$ difference in the initial conditions between these modes. 
 Furthermore, to perturb the translational symmetry along  the $x$ direction, we assume in Eq.~(\ref{eq:translperturb})
 $\epsilon=0.01$, and we take as perturbation frequency $k_0= 0.2 \sqrt{2\omega \mu}$, where the square-root expression is the proper
 frequency of the unperturbed flavor pendulum~\cite{Hannestad:2006nj}. Correspondingly, $P^3_{L,1}(0) =P^3_{R,1}(0) =  \epsilon$ (and analogously for antineutrinos),
while the higher order harmonics are initially vanishing. 
 We follow the evolution of the first  $N=600$ Fourier modes.

%%%%%%%%%%%%%%%%%%%%%%%%%%%%%%%%%%%%%%%%%%%%%%%%%%%%%%%%%%%%%%
\begin{figure}[!t]\centering
\hspace{2.cm}
\includegraphics[angle=0,width=1.\columnwidth]{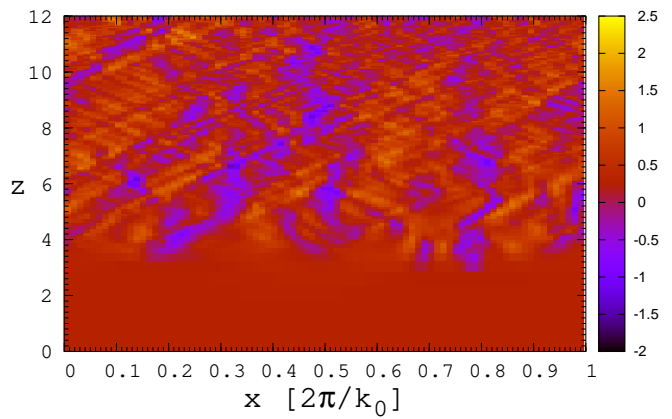}
\vspace{-0.5cm}
\caption{Map of the lepton number ${\textrm L}_0$
 in the $x$-$z$ plane.
\label{lepton}}
\end{figure}
%%%%%%%%%%%%%%%%%%%%%%%%%%%%%%%%%%%%%%%%%%%%%%%%%%%%%%%%%%%%%

In Fig.~1 we show the flavor evolution of $\nu_e$ flavor content $\varrho_{ee}(x,z)$, in the $x$-$z$ plane summed over the $L$ and $R$ contributions and normalized
to $N_{R,L}=2$.
We see that till $z\simeq 2.5$ the flavor evolution shows a translational symmetry,  being  uniform in $x$ direction. The 
flavor content presents the known pendular nutations in $z$ direction, observed in the one dimensional evolution~\cite{Raffelt:2013isa}. Contrarily to what previously 
assumed,
this coherent behavior is not stable.
Indeed, at  $z\simeq 2.5$ the translational symmetry is broken. Then,  $\varrho_{ee}$  develops large variations in the  $x$ direction at length scales that becomes smaller and smaller 
increasing the distance from the boundary.
%, since higher modes ${\sf P}_{(L,R),n}$ are excited.
Furthermore, also the $L \leftrightarrow R$ symmetry is broken, since ${P}^3_L (x,z)$ and 
${P}^3_R (x,z)$  present large differences (not shown), in agreement with the stability analysis of~\cite{Duan:2014gfa}.
The two-dimensional map of the function $\varrho_{ee}(x,z)$ further  clarifies the effect of the breaking of   the translational symmetry.
Till the symmetry is unbroken all the neutrinos oscillate in phase and  the surfaces 
of equal phase are planes parallel to the radiating surface at $z=0$. Then,
when the instabilities develop 
 these planes of common phase are  broken and the coherent behavior of the oscillations is lost.

In Fig.~2 we represent the lepton density ${\textrm L}_0$ of Eq.~(\ref{eq:lepton0}) in the $x$-$z$ plane.
Notice that as soon as the translational instability develops, lepton number shows a non trivial domain structure and that self-induced conversions lead to  
large space variations of the initial  asymmetry $\alpha$.

In order to understand the origin of this flavor dynamics, in Fig.~3 we show a contour plot representing the evolution of the different Fourier modes $|{\sf P}_{R,n}(z)|$ (in logarithmic scale) in the plane
of $n$-$z$. 
The behavior of $|{\sf P}_{L,n}(z)|$ would be similar (not  shown). 
We realize that the breaking of the translational symmetry corresponds to 
the growth of the $n>0$ modes occurring at $z>2$.  
This dynamics can be seen as a cascade process in the Fourier space, 
where a flavor wave caused by the flavor pendulum diffuses to higher harmonics (i.e. to smaller scales) 
as soon as the Fourier modes are excited by the non-linear interactions
between the different modes.
Note the analogy of this process with the multi-angle decoherence
associated with a diffusion of excitations in the multipole space~\cite{Raffelt:2007yz}.
Correspondingly, in the flavor evolution one observes the developments of spatial variations in the $x$ directions at smaller
and smaller scales. 
 In the example we are studying,  at $z=12$ about the first 300 harmonics are significantly excited. Indeed, the number of 
 harmonics that one follows determine the range of validity of the numerical simulation. We checked that the number of excited
harmonics is sensitive
 to   the neutrino-neutrino interaction potential $\mu$, since 
  this factor determines the strength of the terms  responsible for the growth of the modes  in the second line of Eq.~(\ref{eq:eompert}).
The growth of the harmonics is also enhanced   with the initial flavor asymmetry $\alpha$. 
Indeed, in the sum at right-hand-side of  Eq.~(\ref{eq:eompert}) the term ${\sf D}_{(L,R),0}$ increases with the initial flavor asymmetry, 
pumping the higher order harmonics. 
We comment that in realistic cases 
(e.g. for supernova neutrinos)
 one typically has  a declining neutrino potential $\mu$. In this situation one would expect a cut-off in the number of excited modes in function of the distance
from the boundary. 

The multi-angle extension of the results discussed here are presented in Appendix A.

%%%%%%%%%%%%%%%%%%%%%%%%%%%%%%%%%%%%%%%%%%%%%%%%%%%%%%%%%%%%%%
\begin{figure}[!t]\centering
%\hspace{-4.cm}
\includegraphics[angle=0,width=0.9\columnwidth]{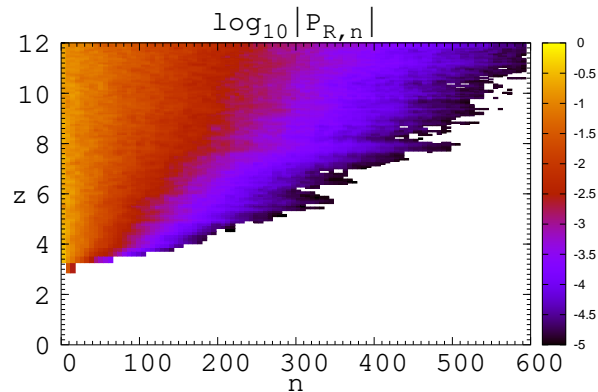}
\vspace{0.cm}
\caption{Contour plot of the first 600 Fourier mode $|{\sf P}_{R,n}|$ (in logarithmic scale) in the plane $n$-$z$.
\label{multipoles}}
\end{figure}
%%%%%%%%%%%%%%%%%%%%%%%%%%%%%%%%%%%%%%%%%%%%%%%%%%%%%%%%%%%%%

The behavior of the neutrino gas in our model has a nice analogy with the transition between laminar and turbulent behavior of a streaming fluid (see, e.g.,~\cite{fluid}).
In this respect, it is useful to define an average neutrino velocity for the neutrino fluid at a point ${\bf x}=(x,z)$.
Considering for example the $\nu_e$ flux, one has
%.........................................................
\begin{equation}
\langle {\hat {\bf v}}_e \rangle_{\bf x} = \frac{\varrho_{ee,L} {\hat {\bf v}}_L + \varrho_{ee,R}{\hat {\bf v}}_R}{\varrho_{ee,L}+\varrho_{ee,R}} \,\ .
\end{equation}
%...........................................................
Till the    $L \leftrightarrow  R$ symmetry is unbroken,we have
$\langle {\hat {\bf v}}_e \rangle_{\bf x} \simeq v_z$. 
Then, when a $L$--$R$ asymmetry with  variations in the $x$ direction is produced, the average velocity starts to acquire a transverse component in the 
$x$ direction.
The ``streamlines'' of the neutrino flux are the solutions of 
%..........................................................................
\begin{equation}
\frac{d{\bf x}}{ds}= \frac{\langle {\hat {\bf v}}_e  \rangle_{\bf x}}{|\langle {\hat {\bf v}}_e \rangle_{\bf x}|}
=\hat{\bf F}_{e,{\bf x}} \,\ ,
\label{eq:stream}
\end{equation}
%............................................................................
where $s$ is a parameter along the line (see~\cite{Dasgupta:2008cu}).

%%%%%%%%%%%%%%%%%%%%%%%%%%%%%%%%%%%%%%%%%%%%%%%%%%%%%%%%%%%%%%
\begin{figure}[!t]\centering
%\hspace{-3.cm}
\includegraphics[angle=0,width=0.8\columnwidth]{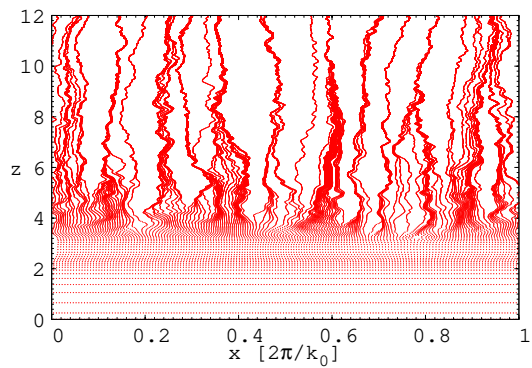}
%\vspace{-1.5cm}
\caption{Streamlines of the $\nu_e$ flux (in vertical direction). 
\label{stream}}
\end{figure}
%%%%%%%%%%%%%%%%%%%%%%%%%%%%%%%%%%%%%%%%%%%%%%%%%%%%%%%%%%%%%

In Fig.~4 we show the  streamlines defined in Eq.~(\ref{eq:stream}) (in vertical direction). One clearly sees that the transition between the coherent to incoherent flavor behavior observed in Fig.~1 corresponds
to the  change from a laminar to a turbulent regime.
As soon as the translational symmetry is broken, 
the streamlines become irregular and are no longer parallel to 
the $z$ directions. Moreover, they exhibit large variations, as in the turbulent  motion in a fluid, and 
 tend to converge in preferred
directions. 

The behavior we found has a fascinating similarity with the non-linear instabilities of fluid flows, described by the Navier-Stokes
equation, see e.g. ~\cite{lunshin,fluid2}, as shown in the Appendix B.

\section{Conclusions} 
%%%%%%%%%%%%%%%%%%%%%%%%%%%%%%%%%%%
We presented the results of  a study of self-induced flavor conversions in a simple  two-dimensional model.  
As predicted from the stability analysis performed in~\cite{Duan:2014gfa}, we find that the self-interacting neutrino gas can break
the spatial symmetries of the initial conditions and the neutrino flavor content achieves large space variations. This implies
that the coherent behavior of the neutrino gas is unstable under small spatial inhomogeneities. 
This process also leads to the formation of domains of different net lepton number ${\textrm L}_0$. 

There is a close analogy between the symmetries of the planar two-dimensional model considered in our work and the
spherical 
SN case. Indeed, the translational symmetry in our model corresponds to the spherical symmetry of the bulb-model 
and the $L$-$R$ symmetry is equivalent to the axial symmetry in the spherical case.
Nevertheless, 
our toy model is much simpler than any realistic SN neutrino case, since we are neglecting multi-angle effects, continuous 
energy spectra, ordinary matter effects and declining neutrino densities. All these effects would add additional complications and numerical challenges. However, if our results would apply also to the SN case this would radically change the current description of the  self-induced flavor conversions,
and would have interesting phenomenological consequences, like e.g., the generation of a self-induced  direction-dependent asymmetry in
the  lepton number flux. 
These intriguing possibilities call for  further efforts to  go beyond the bulb model. This challenge would be  quite demanding in terms of new computig time and/or novel approaches.

%%%%%%%%%%%%%%%%%%%%%%%%%%%%%%%% 
\section*{Acknowledgements} %%%%%%%%%%%%%%%%%%%%%%%%%%%%%%%%%%%%%%%%%%%%%%%%
%%%%%%%%%%%%%%%%%%%%%%%%%%%%%%%%%%%%%%%%%%%%%%%%%%%%%%%%%%%%%%%%%%%%%%
A.M. acknowledges useful discussions with Eligio Lisi, Antonio Marrone and Georg Raffelt. 
 A.M. also thanks Lun-Shin Yao for useful discussions on non-linear instabilities in fluids.
The work of  A.M. was supported by the German Science Foundation (DFG) within the Collaborative Research Center 676 ``Particles, Strings and the Early Universe''
in Germany, and by 
 the Italian Ministero dell'Istruzione, Universit\`a e Ricerca (MIUR) and Istituto Nazionale
di Fisica Nucleare (INFN) through the ``Theoretical Astroparticle Physics'' projects in Italy.
 G.M. is supported by INFN I.S. TASP. 
N.S. acknowledges support from
the European Union FP7 ITN INVISIBLES (Marie Curie
Actions, PITN- GA-2011- 289442).

%%%%%%%%%%%%%%%%%%%%%%%%%%%%%%%%%%%%%%%%%%%%%%%%%%%%%%%%%%%%5
\section*{ Appendix A. Multi-angle case}
%%%%%%%%%%%%%%%%%%%%%%%%%%%%%%%%%%%%%%%%%%%%%%%%%%%%%%%%%%%%%%%%%%%

In the model considered in our work
we have taken    neutrinos to be emitted from the boundary with only two emission modes $L$ and $R$. In this Section
we generalize the model to a multi-angle case. 
As will will see the flavor evolution present substantial differences already without the breaking of the translational symmetry.
Indeed, in the single-angle case one would observe periodic flavor conversions, while in the multi-angle case the emission from an infinite
plane would lead to flavor decoherence. 
Therefore, in comparison with the flavor evolution of SN neutrinos, our single-angle model has to be taken as representative of a case in which 
flavor conversions exhibits a ``quasi-single angle" behavior~\cite{EstebanPretel:2007ec}, while the multi-angle case would represent a realization of the
multi-angle decoherence~\cite{Raffelt:2007yz}.
As we will see, the effect of the breaking of translational symmetry in the two cases produces remarkable differences. 
Indeed,  the large fluctuations of the flavor content found in the single-angle case are smeared-out by the multi-angle effects.

%%%%%%%%%%%%%%%%%%%%%%%%%%%%%%%%%%%%%%%%%%%%%%%%%%%%%%%%%%%%5
\subsection*{Equations of motion}
%%%%%%%%%%%%%%%%%%%%%%%%%%%%%%%%%%%%%%%%%%%%%%%%%%%%%%%%%%%%%%%%%%

With respect to the model considered before we assume 
 $\alpha=1,\ldots N$ neutrino emission modes labeled  in terms of the velocities 
$ {\hat {\bf v}}_{\alpha} = (v_{x,\alpha}, 0,v_{z,\alpha})$, whose components are
$v_{x,\alpha}= \cos \vartheta _{\alpha}$ and $v_{z,\alpha}= \sin \vartheta_{\alpha}$ where
the emission angles 
$\vartheta_{\alpha} \in [0, \pi]$. 
The equations of motion for the polarization vectors  of the $N$ emission modes
are
%..........................................................
\begin{eqnarray}
 {\hat {\bf v}}_{\alpha} \cdot \nabla_{\bf x} {\sf P}_\alpha(x,z)  &=& [+\omega {\sf B} \nonumber \\ 
 &+& \mu  \sum_{\beta=1}^N  (1- {\hat {\bf v}}_{\alpha} \cdot  {\hat {\bf v}}_{\beta})
{\sf D}_{\beta}] \times {\sf P}_{\alpha}(x,z) 
, \nonumber \\
  {\hat {\bf v}}_{\alpha} \cdot \nabla_{\bf x} {\overline{\sf P}}_\alpha(x,z)  &=& 
[-\omega {\sf B}  \nonumber \\
 &+& \mu  \sum_{\beta=1}^N  (1- {\hat {\bf v}}_{\alpha} \cdot  {\hat {\bf v}}_{\beta})
{\sf D}_{\beta}] \times {\overline{\sf P}}_{\alpha}(x,z) \,\ . \nonumber
\label{eq:evolspacemulti}
\end{eqnarray}
%.............................................................
The lepton number in this case is defined as
%...................................................
\begin{eqnarray}
{\textrm L}_0 &=& \sum_{\alpha=1}^N {\sf D}_{\alpha} \cdot{\sf B} \,\ , \label{eq:lepton0} \\
{\bf L} &=& \sum_{\alpha=1}^N  {\hat {\bf v}}_{\alpha} ({\sf D}_{\alpha} \cdot{\sf B}) \,\ ,
\label{eq:lepton}
\end{eqnarray}
%....................................................
where 
${\sf D}_{\alpha} \cdot{\sf B} \simeq P^3_{\alpha} -{\overline P}^3_{\alpha}$.  
The equations for the different Fourier modes [as in Eq.~(11) of the main text]
are
%................................................................
\begin{eqnarray}
v_{z,\alpha} \frac{d}{dz} {\sf P}_{\alpha,n}(z) &=& -i v_{{\alpha},x} k_n {\sf P}_{\alpha,n} + \omega {\sf B}\times {\sf P}_{\alpha,n}  \nonumber \\
&+& \mu \sum_{j=-\infty}^{+\infty}  \sum_{\beta=1}^N
(1-{\hat{\bf v}}_{\alpha} \cdot {\hat{\bf v}}_{\beta})
 {\sf D}_{\beta,{n-j}}\times {\sf P}_{\alpha,j}  \,\ . \nonumber \\
\label{eq:eompert}
\end{eqnarray}
%.....................................................................

%%%%%%%%%%%%%%%%%%%%%%%%%%%%%%%%%%%%%%%%%%%%%%%%%%%%%%%%%%%%%%%%%%%%%%%%%%%

%.......................................................................
\subsection*{Numerical examples}
%......................................................................

\begin{figure}[!t]\centering
\hspace{-1.cm}
\includegraphics[angle=0,width=0.9 \columnwidth]{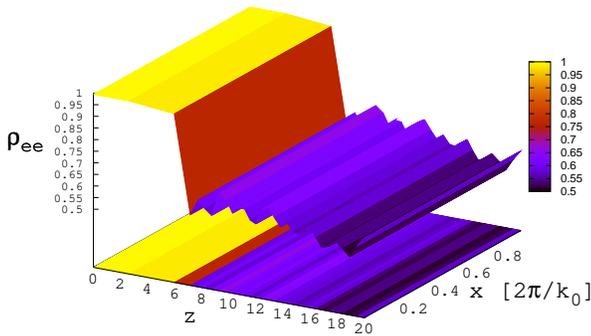}
%\vspace{-3.cm}
\caption{Multi-angle flavor evolution: 
Two-dimensional evolution of the ${\bar\nu}_e$ flavor content ${\bar\varrho}_{ee}$ in the $x$-$z$ plane, and its map  on the bottom plane
assuming the translational invariance along the $x$-direction.
\label{fig9a}}
\end{figure}

In this Section we show 
how the presence of multi-angle effects modifies the previous single-angle results. This example is intended to capture
a fundamental feature of the non-isotropic  neutrino emission from a supernova core. 
For this reason we consider the same case of the work, assuming that each point along the boundary
emits $N=200$ angular modes with emission angles $\vartheta_{\alpha}$  equally distributed in  the range $[0, \pi]$. 
We checked that the behavior  of the flavor conversions in qualitatively similar further increasing the number of emission modes.
In Fig.~\ref{fig9a} we consider a case in which  the translational invariance along the $x$-direction is
assumed. We realize that instead of the periodic flavor conversions observed expected in the single-angle case  one now finds a flavor decoherence
with  ${\bar\varrho}_{ee} (x,z) \simeq 0.5$ as soon as flavor conversions start at $z \gtrsim 6$. This case has been widely studied 
in~\cite{Raffelt:2007yz} where it has been realized that in the presence of a constant neutrino potential, the multi-angle decoherence
is unavoidable in both normal and inverted mass hierarchy.

\begin{figure}[!t]
\centering
\hspace{-1.cm}
\includegraphics[angle=0,width=0.9 \columnwidth]{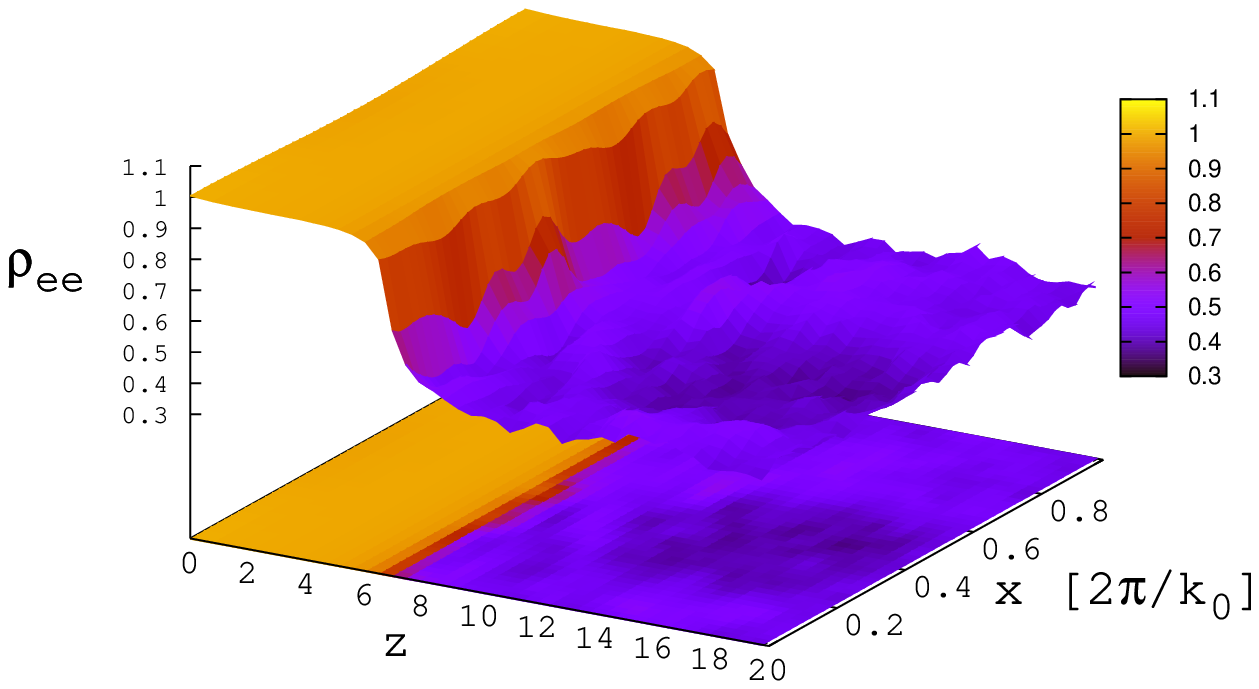}
%\vspace{-3.cm}
\caption{Multi-angle flavor evolution:
Two-dimensional evolution of the ${\bar\nu}_e$ flavor content ${\bar\varrho}_{ee}$ in the $x$-$z$ plane, and its map  on the bottom plane
breaking the translational invariance along the $x$-direction.
\label{fig9}}
\end{figure}

We show now how this result is affected by the breaking of the translational invariance.
At this regard, in Fig.~\ref{fig9} we consider the multi-angle versions of the flavor evolution
shown for the single-angle case in  Fig.~1 of the main text.
We realize  that the map of the ${\bar\varrho}_{ee} (x,z)$ is rather different with respect to the corresponding one in the single
angle case.
In particular, when flavor conversions start they quickly lead to a flavor equilibrium with ${\bar\varrho}_{ee} (x,z) \simeq 0.5$
across the plane.  
The flavor variations along the $x$ direction are smoothed with respect to the single-angle case being at most $\sim 20$\%.
From this figure one realizes that the output of the flavor evolution for a constant $\mu$ is the  multi-angle decoherence, 
as  also found  in the one-dimensional models~\cite{Raffelt:2007yz}.

\begin{figure}[!t]\centering
\hspace{-1.cm}
\includegraphics[angle=0,width=0.9\columnwidth]{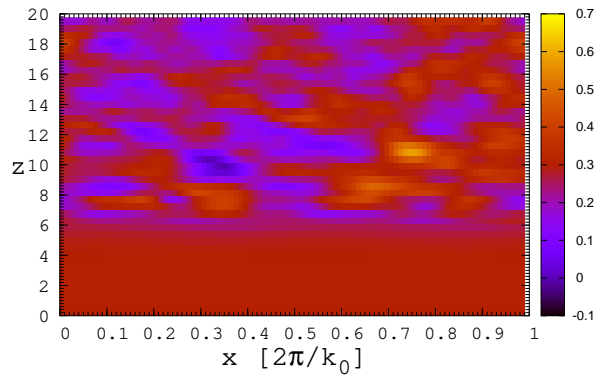}
%\vspace{-2.5cm}
\caption{Multi-angle flavor evolution:  Map of the lepton number ${\textrm L}_0$.
\label{fig15}}
\end{figure}

In Fig.~\ref{fig15} we show the evolution of the lepton numeber ${\textrm L}_0$ 
in the $x$-$z$ plane.
With respect to the corresponding single-angle case [Fig.~2 of the main text] we realize 
that the spatial variations in ${\textrm L}_0$ are $\sim 0.3$. Then, they are strongly reduced
with respect to the single-angle case. However, also in this case we can have the formation
of domain of lepton number with opposite sign.

%%%%%%%%%%%%%%%%%%%%%%%%%%%%%%%%%%%%%%%%%%%%%%%%%%%%%%%%%%%%%%
%%%%%%%%%%%%%%%%%%%%%%%%%%%%%%%%%%%%%%%%%%%%%%%%%%%%%%%%%%%%%

%%%%%%%%%%%%%%%%%%%%%%%%%%%%%%%%%%%%%%%%%%%%%%%%%%%%%%%%%%%%%%
%%%%%%%%%%%%%%%%%%%%%%%%%%%%%%%%%%%%%%%%%%%%%%%%%%%%%%%%%%%%%%%%%%%%%%%%%%%%%%%%%%%%%%%%%%%

%%%%%%%%%%%%%%%%%%%%%%%%%%%%%%%%%%%%%%%%%%%%%%%%%%%%%%%%%%%%%%
%%%%%%%%%%%%%%%%%%%%%%%%%%%%%%%%%%%%%%%%%%%%%%%%%%%%%%%%%%%%%

%%%%%%%%%%%%%%%%%%%%%%%%%%%%%%%%%%%%%%%%%%%%%%%%%%%%%%%%%%%%%%
\begin{figure}[!t]\centering
\hspace{-1.cm}
\includegraphics[angle=0,width=0.9\columnwidth]{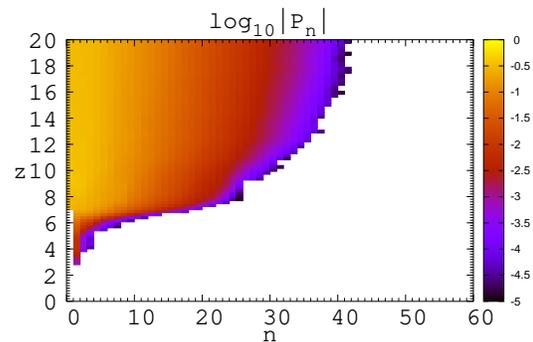}
%\vspace{-3.cm}
\caption{Multi-angle flavor evolution: Contour plot of the first 600 Fourier mode $|{\sf P}_{n}|$ (in logarithmic scale) in the plane $n$-$z$.
\label{fig12}}
\end{figure}
%%%%%%%%%%%%%%%%%%%%%%%%%%%%%%%%%%%%%%%%%%%%%%%%%%%%%%%%%%%%%

Finally, in   Fig.~\ref{fig12} we compare  the evolution of the different Fourier modes $|{\sf P}_{n}(z)|$ (summed over the emission angles) in the plane
of $n$-$z$,  for the case discussed above.
We realize that the number  of excited Fourier modes is strongly reduced (by an order of magnitude) with respect to the 
corresponding single-angle case [see Fig.~3 in the main text]. 
The reason is that when the decoherence quickly starts the length of $|{\sf P}_0|$ is significantly shortened.
As one realizes from the non-linear term at second line in Eq.~(\ref{eq:eompert})
this effect reduces the growth of the higher order harmonics. Moreover, the multi-angle kernel in the higher order terms
leads itself to a suppression of the higher order harmonics.
This reduced growth of the higher order modes causes a strong suppression in the spatial variations of
the flavor content. 

%%%%%%%%%%%%%%%%%%%%%%%%%%%%%%%%%%%%%%%%%%%%%%%%%%%%%%%%%%%%5
\section*{Appendix B. Analogy with non-linear fluid instabilities}
%%%%%%%%%%%%%%%%%%%%%%%%%%%%%%%%%%%%%%%%%%%%%%%%%%%%%%%%%%%%%%%%%%%
As discussed before, the transition of the neutrino gas from a coherent to incoherent behavior  shows an intriguing  analogy  with  a streaming flow changing from  laminar to  turbulent regime. 
In this Section we discuss more closely the physical ground of this  similarity, closely following the presentation of~\cite{fluid,lunshin,fluid2} concerning non-linear instabilities of a  fluid.
The fluid motion is governed by the following equations~\cite{fluid}
%..............................................................
\begin{eqnarray}
\nabla\cdot {\bf u} &=& 0 \,\ , \nonumber \\
\frac{\partial {\bf u}}{\partial t}+ {\bf u} \cdot 
\nabla {\bf u} &=& - \frac{\nabla p}{\rho} + \nu \nabla^2 {\bf u} \,\ ,
\label{eq:navier}
\end{eqnarray}
%...............................................................
where the first equation is the continuity equation, while the 
second are the incompressible Navier-Stokes momentum equations,
where ${\bf u}=(u,v,w)$ are the velocity components in the 
$(x,y,z)$ directions respectively, $p$ is the pressure, $t$ is the time
and $\nu$ the viscosity.
The non-linear term   ${\bf u} \cdot 
\nabla {\bf u}$ is the one responsible for the turbulent behavior of a fluid.
Eq.~(\ref{eq:navier}) admits a steady solution ${\bf u}= U(y), v=w=0$ for properly defined 
boundary conditions. 
In order to study the  stability of this steady solution, one superimposes a  perturbation  
on the basic flow, obtaining as disturbed velocity
%...................................................
\begin{equation}
{\bf u} = (u,v,w)= (U(y) + u^\prime, v^\prime, w^\prime) \,\ .
\label{eq:disturb}
\end{equation}
%.....................................................
Substituting Eq.~(\ref{eq:disturb}) into Eq.~(\ref{eq:navier}) one obtains
a system of equations for the disturbance.
One can express the disturbance as \emph{ Fourier integral} over all possible continue wavenumbers
%......................................................
\begin{equation}
{\bf u}^{\prime}(x,y,t) = \int_{-\infty}^{+\infty} {\hat {\bf u}} ({\bf k},y,t)e^{i {\bf k}\cdot{\bf x}} d{\bf k} \,\ ,
\end{equation}
%.........................................................
where ${\bf x}=(x,z)$ and ${\bf k}=(k_x,k_z)$. 
One can search for solutions of Eq.~(\ref{eq:navier}) expressing the 
Fourier components as 
%......................................................
\begin{equation}
{\bf u}^{\prime}(x,y,t) = \sum_{m=1}^{\infty} A_m({\bf k},t) {\tilde {\bf U}}_m ({\bf k},y) \,\ ,
 \end{equation}
%......................................................
where ${\tilde{\bf U}}_m$  is the eigenfunction corresponding to the $m$-th eigenfrequency $\omega_m$, and 
$A_m$ is a time-dependent amplitude-density function.
One obtains the following equations for the amplitude-density functions
%................................................
\begin{equation}
\frac{d A_m}{dt}+ i\omega_m A_m = \sum_{m_1=1}^{\infty} \sum_{m_2=2}^\infty I({\bf k},m,m_1,m_2,t) \,\ ,
\label{eq:amplitude}
\end{equation}
%....................................................
where 
\begin{eqnarray}
& &I({\bf k},m,m_1,m_2,t)= \nonumber \\
& &\int_{-\infty}^{\infty} b({\bf  k},{\bf k}_1,m,m_1,m_2) A_{m_1}({\bf k}_1,t)
A_{m_2}({\bf k}-{\bf k}_1,t) d{\bf k}_1 \,\ . \nonumber \\
\label{eq:nonlinear}
\end{eqnarray} 
 The amplitude of the waves is determined
by the linear and non-linear terms.
 Laminar flow contains only one wave, i.e. the mean flow. 
The linear term in Eq.~(\ref{eq:amplitude})
represents the mean-flow convection 
of the velocity disturbance.
The non-linear term [Eq.~(\ref{eq:nonlinear})]  involves energy 
transfer among different  waves satisfying resonance conditions so that the excited wavenumbers are discrete.
All waves can resonantly interact with the mean flow. 

After this discussion about non-linear behavior of fluid it is apparent that the strategy used by us 
to follow the evolution of the  polarization vectors ${\sf P}$ in flavor space is analogous to 
the one used to characterize the  velocity field ${\bf v}$ of a flow. 
Indeed, in both the cases the use of Fourier transform allows one to pass from a partial differential equation problem to a tower
of ordinary differential equations for the different Fourier modes.  
In both cases one is interested to study the evolution of disturbances with respect to a steady one-dimensional 
behavior. 
The amplitude of the excited modes is determined by linear and non-linear term.
Neglecting the non-linear term one can apply a stability analysis to the linearized
 equations of motion as done in~\cite{Duan:2014gfa}
 for the flavor problem and in~\cite{lunshin} for the Navier-Stokes equations.
What we have done in the current work is to study the flavor dynamics in the non-linear regime, characterized 
by a convolution term among different harmonics. This term [Eq.~(11) of the main text] has a structure similar to 
Eq.~(\ref{eq:nonlinear}) for the case of a fluid.
The similar mathematical form of the problem explains the intriguing analogy among the 
transition between laminar and turbulent regime of a fluid and between coherent and incoherent behavior 
of the neutrino gas. Indeed as for a laminar fluid only one wave exists, for the coherent flavor conversions only the $n=0$
Fourier mode evolves. 
Moreover, the interaction among the different harmonics that are rapidly excited breaks the coherent behavior of the flavor conversions
for neutrinos and leads to the passage into the turbulent regime for a fluid.

We believe that this nice analogy would be further developed to get a deeper understanding of the neutrino flavor evolution.
Moreover, one would also benefit of the huge literature developed to study the non-linear Navier-Stokes equations.

\end{document}